\documentclass{article}
\usepackage{spconf,amsmath,graphicx,hyperref}

\usepackage{cite}
\usepackage{amsmath,amssymb,amsfonts}
\usepackage{algorithmic}
\usepackage{graphicx}
\usepackage{textcomp}
\usepackage{xcolor}
\usepackage{booktabs}
\usepackage{multirow}
\usepackage{enumitem}

\newcommand{\remark}[1]{}
\newenvironment{hide}[1]{\remark{#1}}{}

\newcommand{\full}[1]{#1}                 
\newcommand{\short}[1]{}              

\begin{document}

\title{Etude: Piano Cover Generation with a Three-Stage Approach --- Extract, strucTUralize, and DEcode}

\name{Tse-Yang Chen and Yuh-Jzer Joung}
\full{\address{Dept.~of Information Management, National Taiwan University, Taipei, Taiwan}}
\short{\address{National Taiwan University, Taipei, Taiwan}}
\begin{hide}{
\author{\IEEEauthorblockN{1\textsuperscript{st} Tse-Yang Chen}
\IEEEauthorblockA{\textit{Dept.~of Information Management} \\
\textit{National Taiwan University}\\
Taipei, Taiwan \\
r12725050@ntu.edu.tw}
\and
\IEEEauthorblockN{2\textsuperscript{nd} Yuh-Jzer Joung}
\IEEEauthorblockA{\textit{Dept.~of Information Management} \\
\textit{National Taiwan University}\\
Taipei, Taiwan \\
joung@ntu.edu.tw}
}
}\end{hide}
\maketitle

\begin{abstract}
Piano cover generation aims to automatically transform a pop song into a piano arrangement.
While numerous deep learning approaches have been proposed, existing models often fail to maintain structural consistency with the original song, likely due to the absence of beat-aware mechanisms or the difficulty of modeling complex rhythmic patterns. Rhythmic information is crucial, as it defines structural similarity (e.g., tempo, BPM) and directly impacts the overall quality of the generated music.
In this paper, we introduce  \textbf{Etude}, a three-stage architecture consisting of \textbf{E}xtract, struc\textbf{TU}ralize, and \textbf{DE}code stages. By pre-extracting rhythmic information and applying a novel, simplified REMI-based tokenization, our model produces covers that preserve proper song structure, enhance fluency and musical dynamics, and support highly controllable generation through style injection. Subjective evaluations with human listeners show that Etude substantially outperforms prior models, achieving a quality level comparable to that of human composers.
\end{abstract}


\short{\renewcommand{\baselinestretch}{0.9}}



\section{Introduction}\label{sec:introduction}

Piano cover generation—the artistic rearrangement of a song into a piano performance—is a popular form of musical expression that thrives on social media and has grown into a commercially viable field\full{, with applications ranging from enabling learners to obtain customized sheet music to assisting creators in producing arrangements for streaming or distribution}.
Yet producing a high-quality piano cover remains challenging, requiring both analysis of the original song’s musical features (e.g., harmony and rhythm) and sufficient music theory knowledge to create a coherent, aesthetically pleasing arrangement.


Numerous studies \cite{pop2piano,picogen,picogen2,amt-apc,music2midi} have explored Automatic Piano Cover Generation (APCG) with deep learning. These methods often emulate the two-step creative process of human composers: first analyzing music-theoretical features such as harmony, melody, and rhythm, and then recombining them to synthesize a new interpretation within the piano’s 88-key range. Nevertheless, a substantial quality gap remains between model-generated and human-composed covers, as reflected in subjective evaluation scores, suggesting that the task’s complexity and nuance have yet to be fully captured.

\full{
\begin{figure}
  \centering
  \includegraphics[width=0.9\linewidth]{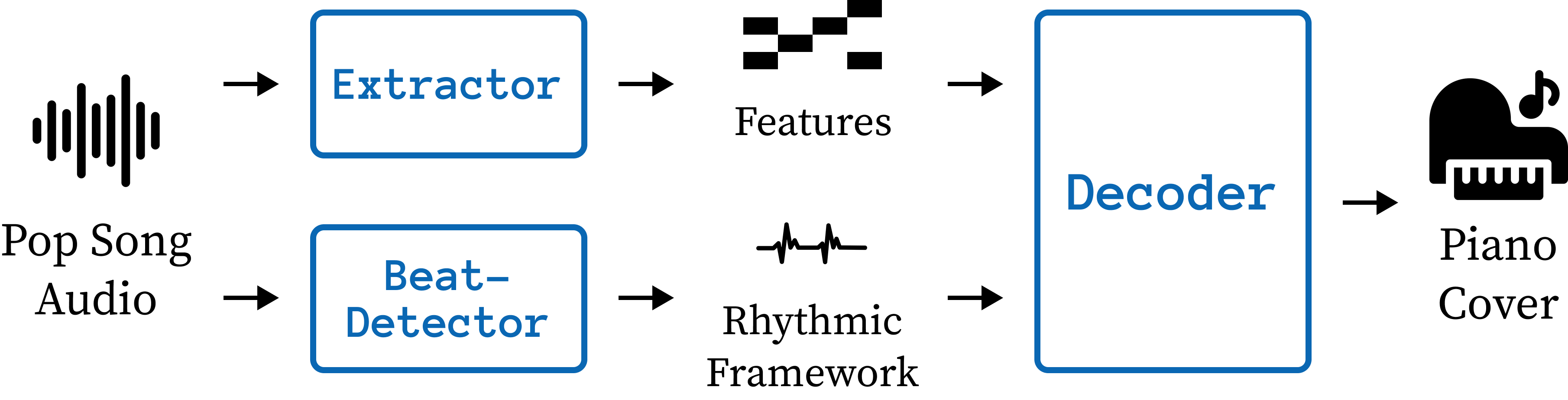}
 \full{ \caption{The architecture of our proposed Etude framework, which comprises three main components: an Extractor for the extract stage, a Beat-Detector for the structuralize stage, and a Decoder for the decode stage.}}
\short{\caption{The architecture of our proposed Etude framework.}}
  \label{fig:architecture}
\end{figure}
}


A closer examination reveals a key source of this quality gap: prior models struggle to capture a song’s rhythmic and structural framework. Their outputs frequently lack consistency in rhythm, tempo, or time signature—with some models \cite{pop2piano,amt-apc,music2midi} relying on time-based rather than beat-based frameworks, resulting in off-beat notes and degraded quality. We attribute this shortcoming to an architectural limitation: because beat tracking is notoriously difficult in music information retrieval (MIR), asking models to implicitly learn a song’s structure while simultaneously extracting other musical features creates a compounded challenge that undermines their effectiveness.

\begin{figure*}[t]
  \centering
\full{  \includegraphics[width=0.85\textwidth]{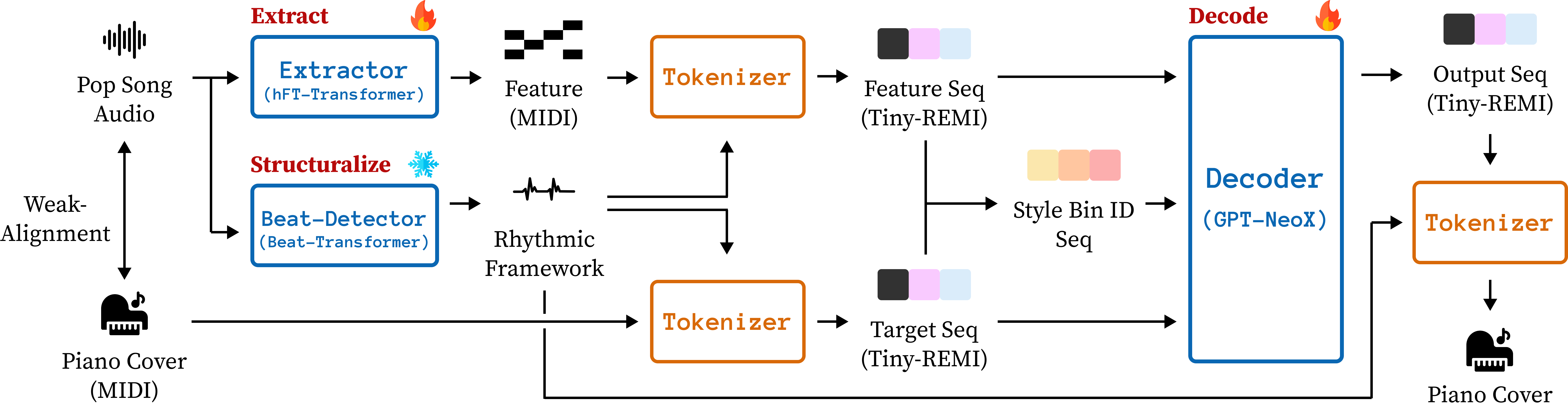}}
\short{  \includegraphics[width=0.75\textwidth]{figures/full-model.png}}
\caption{
\full{
The overall architecture of \textbf{Etude}, where an \textbf{Extractor} derives harmonic features, a pre-trained \textbf{Beat-Detector} provides the rhythmic framework for tokenization and de-tokenization, and a \textbf{Decoder} receives these inputs along with style prompts. During training, the lower piano cover branch provides ground-truth target sequences ($Y$) aligned via weak-alignment. During inference, the lower branch content is replaced by autoregressively generated outputs: the Decoder generates each bar $Y_i$ from the corresponding feature bar $X_i$ and previous context. The Extractor and Decoder are trained separately.
}
\short{The overall architecture of \textbf{Etude}.}
}
 \label{fig:full-model}
\end{figure*}

To address this issue, we introduce \textbf{Etude}, a novel architecture that builds upon the two-stage concept of PiCoGen \cite{picogen,picogen2}, but further modularizes the task by explicitly separating the beat detection process. \full{ As illustrated in Figure~\ref{fig:architecture}, the name}\short{The name} \textbf{Etude} 
reflects the model's three distinct stages: the \textbf{E}xtract stage for extracting music-theoretical features, followed by the struc\textbf{TU}ralize stage for deriving the rhythmic framework, and concluding with the \textbf{DE}code stage where all features are combined to generate the final output\short{ (see Figure~\ref{fig:full-model})}. The key advantages of our approach are twofold. First, by explicitly extracting the rhythmic framework, our model guarantees structural consistency with the original song, which significantly enhances overall quality and musicality. Second, our architecture allows for the injection of specified styles, a feature that increases both the controllability of the output and its practical application value.\full{

In summary, by disentangling the learning of structure from the generation of notes, Etude produces structurally sound, stylistically controllable, and high-quality piano covers, demonstrating state-of-the-art performance in extensive evaluations where it substantially outperformed previous methods and achieved a quality closer to human-composed arrangements.} 
All code and audio demonstrations, along with a full version of the manuscript, are available on our project page.\footnote{https://xiugapurin.github.io/Etude/}

\full{
\section{Background}\label{sec:background}

The task of Automatic Piano Cover Generation (APCG) presents several unique challenges for deep learning models. Among these, a primary requirement is the ability to comprehend long-term musical context spanning multiple measures\cite{music-transfomer}, which hinges on the development of an effective symbolic music representation\cite{symbolic-music-survey}. Furthermore, to enhance the model's practical utility and user control, incorporating mechanisms from conditional music style transfer is essential\cite{midi-vae}.

Modern symbolic music generation is fundamentally built on event-based representations. An early and intuitive approach is the piano roll, which encodes music as a 2D pitch–time matrix and is well-suited to image-based models such as CNNs~\cite{midi-net,muse-gan,lead-sheet-gen}. However, it lacks explicit note-off information, making it difficult to distinguish between long sustained notes and repeated short notes \cite{symbolic-music-survey}. Event-based sequences, by contrast, use MIDI-like tokens (e.g., `Note On’ and `Note Off’) that are more explicit but come with limitations: they struggle to represent rhythmic structure clearly, are inefficient to model, and may produce artifacts such as hanging notes. A major advancement came with REMI~\cite{remi}, which introduced \textit{Bar} and \textit{Position} tokens to provide a metrical grid for rhythm. While effective, REMI often generates very long sequences, posing challenges for Transformer-based models. To address this, the Compound Word (CP) representation \cite{cp-word} was proposed, grouping related events into  `words’ to shorten sequence length. These beat-aware and compact representations have proven crucial for enabling state-of-the-art models like the Transformer to capture long-range musical dependencies.


Beyond representation, controllable music style transfer has emerged as a central research focus. Early work (e.g., \cite{transfer-style-of-homophonic-music}) adapted steerable architectures such as DeepBach \cite{deepbach} for homophonic accompaniment style transfer. A persistent challenge is the scarcity of aligned data, motivating predominantly unsupervised approaches. VAE-based methods, such as MIDI-VAE \cite{midi-vae}, addressed this by disentangling style in a shared latent space with classifier guidance, while GAN-based frameworks, e.g., CycleGAN \cite{cyclegan-style-transfer}, leveraged image processing techniques for symbolic genre transfer. More recent research emphasizes attribute-based control, with systems like MuseMorphose \cite{musemorphose}, inspired by Music FaderNets \cite{music-fadernets}, enabling fine-grained manipulation of perceptual dimensions such as rhythmic intensity and polyphonic density.


Although these techniques provide a strong foundation, their application to APCG exposes several key challenges. PiCoGen \cite{picogen,picogen2} established the two-stage paradigm: extracting core features from a song into a simplified intermediate representation and then generates a piano cover. However, 
this intermediate step often acts as an information bottleneck, losing much of the original audio's nuance. 
Data alignment further compounds the problem: strong-alignment methods \cite{pop2piano} often induce audio artifacts, whereas weak-alignment approaches \cite{picogen2} remain limited by inaccuracies in beat-tracking.  In contrast, models like AMT-APC\cite{amt-apc} utilize a more direct transcription approach that captures richer features but often at the cost of explicit high-level structural modeling.


A common limitation across these methodologies is the lack of \textbf{structural consistency}. Data artifacts, information bottlenecks, and the inherent difficulty of implicitly modeling rhythm all contribute to failures in preserving the original song’s rhythmic framework—tempo, bar, and beat alignment. This gap in structural integrity forms the central motivation for our work.
}

\section{Methodology}\label{sec:methodology}
\full{
Our proposed framework, Etude, is designed to address the key challenges in APCG, particularly in ensuring structural consistency and enabling stylistic control. }
The architecture, illustrated in Figure \ref{fig:full-model}, modularizes the generation process into three distinct stages: Extract, Structuralize, and Decode. \full{ This section details the motivation and implementation of each stage.} 

\subsection{Extract Stage}
The goal of the Extract stage is to overcome the information bottleneck found in prior two-stage models\cite{picogen,picogen2}, which often use simplified lead sheet\cite{sheet-sage} representations that lose significant musical nuance. To address this, we produce a rich and dense feature representation that captures all potentially salient events from the source audio. Our Extractor's architecture is adapted from AMT-APC\cite{amt-apc}, 
which is a piano cover generation model fine-tuned from the hFT-Transformer\cite{hft-transformer} for automatic  music transcription.
We use the source audio as input but modify the model's loss sampling parameter ($\theta^{\text{matrix}}$) to encourage the transcription of a dense map of musical events rather than a sparse, playable arrangement. The final output is a MIDI-like feature sequence that provides a comprehensive and detailed foundation for the Decoder.

\subsection{Structuralize Stage}
The Structuralize stage is central to ensuring structural correctness. Running in parallel with the Extractor, it employs a pre-trained Beat-Transformer\cite{beat-transformer} to analyze the source audio and extract the precise timings of all beats and downbeats. This information forms a definitive rhythmic framework $F_{beat}$, containing tempo, time signature, and measure boundaries, that serves as an immutable structural ground truth for the entire process. This framework underpins our decoupled approach: it provides all necessary metrical information for the tokenization of symbolic data and the final reconstruction of a structurally coherent MIDI file from the model's output (see Section \ref{sec:data-representation} for details on how $F_{beat}$ is used in tokenization and de-tokenization).

\subsection{Decode Stage}
The Decode stage employs a Transformer-based model to synthesize the final piano cover. Its primary goal is to translate the dense feature sequence ($X$) from the Extractor into the target symbolic piano cover sequence ($Y$), conditioned on a set of style attributes that guide the arrangement process.

To teach the model this complex translation task, we employ a bar-wise mix\cite{compose-embellish,picogen,picogen2} training strategy. For each song, the feature sequence $X$ and target sequence $Y$ are segmented by measure and interleaved into a single sequence of the form $[X_1,Y_1,X_2,Y_2,...]$. To distinguish between the two sources, each token is accompanied by a parallel Class ID (\texttt{SRC} for feature tokens, \texttt{TGT} for target tokens). The model is trained to predict the target bar $Y_i$, using its corresponding feature bar $X_i$ and the context of the preceding four mixed bars. This four-bar context window is designed to capture the common phrasing structures found in popular music, thereby enhancing the musical continuity of the output.

During inference, the bar-wise mixing structure is maintained, but the target sequences Y come from the model's own generations rather than ground-truth piano covers. Specifically, the generation proceeds bar-by-bar: for each bar $i$, the Decoder receives the current feature bar $X_i$ along with up to four preceding $(X, Y)$ pairs as context. 
The model then autoregressively generates tokens for $Y_i$ until a bar-end token is produced. The completed pair $(X_i, Y_i)$ is appended to the context window for generating the next bar. This approach ensures that the model leverages its own outputs to maintain long-term musical coherence during generation.

To enable controllable generation and address the one-to-many nature of the arrangement task, we introduce \textbf{style vectors} that explicitly describe the relationship between a feature bar ($X_i$) and its corresponding target bar ($Y_i$). We designed three relative style attributes inspired by prior work\cite{musemorphose,music-fadernets} on controllable generation:

\short{ \begin{itemize}[noitemsep,nolistsep]}
\full{\begin{itemize}}
\item \textbf{Relative Polyphony}: 
change in the average number of notes per unique time event. It reflects a shift in harmonic texture; for example, a high value indicates a transformation towards denser block chords, while a low value suggests a shift towards sparser textures like arpeggios.
\item \textbf{Relative Rhythmic Intensity}: 
change in the density of rhythmic events over time. A higher value results in a more complex and rhythmically active phrase.
\item \textbf{Relative Note Sustain}: 
change in the average note duration, controlling the output's articulation character from legato (high) to staccato (low).
\end{itemize}

These relative attributes are calculated for every bar pair in the training set and discretized into three bins {0, 1, 2} (low, medium, high)  based on their statistical distribution. During training and inference, each token in a bar-pair sequence is assigned the three corresponding style bin IDs for that bar. These IDs are then passed through separate embedding layers, concatenated, and projected by a linear layer to form a single style embedding vector. This style embedding is added to the token and class ID embeddings to form the final input to the Transformer Decoder.\full{ This mechanism effectively conditions the entire generation process on the desired stylistic transformation, allowing for fine-grained, bar-level control over the final output's characteristics.}

\begin{figure*}[t]
  \centering
  \includegraphics[width=0.7\textwidth]{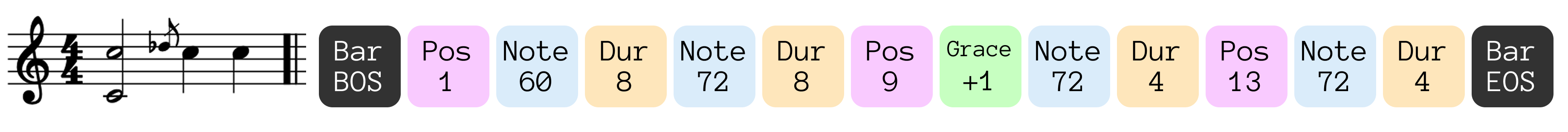}
  \caption{An example of Tiny-REMI tokenization, where the sequence is delimited by \texttt{Bar} [BOS] and \texttt{Bar} [EOS] tokens, and each note event group (\texttt{Note} + \texttt{Dur}) is preceded by a \texttt{Pos} token marking its relative position.}
  \label{fig:tiny-remi}
\end{figure*}

\subsection{Data Representation}\label{sec:data-representation}
The choice of data representation is critical for the success of a symbolic music generation model. While prior works like Pop2Piano\cite{pop2piano} used raw MIDI event sequences, which are difficult for sequence models to learn, subsequent research\cite{picogen,picogen2} has gravitated towards more structured representations\cite{remi,cp-word}. However, the standard REMI tokenizer, designed for general-purpose music generation, includes tokens for \texttt{Tempo}, \texttt{Chord}, and \texttt{Velocity}. We argue that for the APCG task, forcing the Decoder to learn these attributes introduces unnecessary complexity. For instance, tempo information is already explicitly handled by our pre-extracted rhythmic framework ($F_{\text{beat}}$), while other attributes like chords and velocity are not essential for learning the core structural translation task that is our focus.

To address this, we designed \textbf{Tiny-REMI}, a minimal and efficient token set tailored specifically for the APCG task. Our representation consists of five token classes. \texttt{Bar} tokens ([BOS], [EOS]) are special tokens used to mark the beginning and end of each measure. \texttt{Pos} is a metric-related token indicating the 16th-note offset within a measure. The remaining tokens are note-related: \texttt{Note} for note pitches (88 values), \texttt{Duration} for note durations (10 common values based on 16th-note units), and \texttt{Grace} for representing grace notes ($\pm 1$ semitone). Figure \ref{fig:tiny-remi} provides an example of a musical measure tokenized in this format. By removing redundant tokens, this minimalist vocabulary simplifies the learning task for the Decoder, allowing it to focus solely on the relationship between note events.

The encoding and decoding processes both rely on $F_{\text{beat}}$. To encode a MIDI file, each note's absolute onset time is quantized and mapped to a relative \texttt{Pos} token within its corresponding measure. The decoding process is the reverse: the Decoder's output sequence, which contains only relative positional information, is rendered back into a standard MIDI file by using $F_{\text{beat}}$ to restore absolute timings, tempo, and meter for each measure. This decoupled approach ensures that the generated output is not only musically coherent but also structurally sound and easily convertible to standard musical formats like MIDI or MusicXML.

\section{Experiments}\label{sec:experiments}

\begin{table*}[ht]
\caption{Comprehensive evaluation results of human composers, our proposed Etude models, and other baselines on both objective and subjective metrics. For objective metrics, we also show the delta ($\Delta$) from the human performance. For subjective metrics, higher is better ($\uparrow$).}
\centering
\setlength{\tabcolsep}{5pt}
\begin{tabular}{l ccc cccc}
\toprule
\textbf{Model} & \multicolumn{3}{c}{\textbf{Objective Evaluation}} & \multicolumn{4}{c}{\textbf{Subjective Evaluation} ($\in$ [1, 5])}  \\
\cmidrule(lr){2-4} \cmidrule(lr){5-8}
& \textbf{WPD} & \textbf{RGC} & \textbf{IPE} & \textbf{SI $\uparrow$} & \textbf{FL $\uparrow$} & \textbf{DE $\uparrow$} & \textbf{OVL $\uparrow$} \\
\midrule
Human & 0.49 & 0.042 & 10.13 & 3.75 $\pm$ 1.10 & 4.03 $\pm$ 1.02 & 3.79 $\pm$ 1.06 & 3.92 $\pm$ 0.96 \\
\midrule
Etude - Default & 0.21 {\small($\Delta$ 0.28)} & 0.020 {\small($\Delta$ 0.022)} & 9.02 {\small($\Delta$ 1.11)} & 3.16 $\pm$ 1.07 & \textbf{3.73} $\pm$ 0.98 & 3.46 $\pm$ 1.05 & \textbf{3.50} $\pm$ 0.99 \\
Etude - Prompted & 0.23 \textbf{\small($\Delta$ 0.26)} & 0.026 {\small($\Delta$ 0.016)} & 9.11 {\small($\Delta$ 1.02)} & 3.17 $\pm$ 1.10 & 3.70 $\pm$ 1.05 & \textbf{3.49} $\pm$ 1.06 & 3.46 $\pm$ 1.00 \\
Etude Extractor & 0.12 {\small($\Delta$ 0.37)} & 0.028 \textbf{\small($\Delta$ 0.014)} & 10.62 \textbf{\small($\Delta$ 0.49)} & \textbf{3.41} $\pm$ 1.01 & 3.31 $\pm$ 1.13 & 3.35 $\pm$ 1.03 & 3.33 $\pm$ 1.00 \\
\midrule
PiCoGen2\cite{picogen2} & 1.00 {\small($\Delta$ 0.51)} & 0.059 {\small($\Delta$ 0.017)} & 7.97 {\small($\Delta$ 2.16)} & 2.88 $\pm$ 1.13 & 3.33 $\pm$ 1.12 & 2.73 $\pm$ 1.14 & 2.97 $\pm$ 1.04 \\
AMT-APC\cite{amt-apc} & 0.09 {\small($\Delta$ 0.40)} & 0.114 {\small($\Delta$ 0.072)} & 10.69 {\small($\Delta$ 0.56)} & 2.64 $\pm$ 0.99 & 2.37 $\pm$ 1.11 & 2.71 $\pm$ 1.13 & 2.46 $\pm$ 1.04 \\
Music2MIDI\cite{music2midi} & 0.18 {\small($\Delta$ 0.31)} & 0.160 {\small($\Delta$ 0.118)} & 8.94 {\small($\Delta$ 1.19)} & 2.56 $\pm$ 1.06 & 2.29 $\pm$ 1.13 & 2.24 $\pm$ 1.09 & 2.27 $\pm$ 1.07 \\
\bottomrule
\end{tabular}
\label{tab:evaluation}
\end{table*}

\subsection{Dataset}\label{sec:dataset}

We collected a dataset of approximately 7,700 pop song and piano cover audio pairs, primarily consisting of J-pop and K-pop. To ensure data quality, we filtered out pairs with a length difference greater than 30 seconds or a WP-std\cite{music2midi} greater than 1.0. The remaining pairs were then synchronized using the weakly-alignment method proposed in PiCoGen2\cite{picogen2}.\footnote{\full{Our process modifies the original PiCoGen2 method. }
 Instead of performing beat detection on both audio tracks, we detect beats\cite{beat-transformer} only on the source audio. We then find the corresponding timestamps in the cover audio using the MrMsDTW\cite{mrmsdtw} implementation from the SyncToolbox Python package\cite{sync-toolbox} before aligning at the \textbf{measure} level.} Our final training dataset consists of 4,752 pairs, totaling approximately 500 hours of audio. For evaluation, we curated a separate test set of 100 songs not seen during training, evenly distributed across four genres: C-pop, J-pop, K-pop, and Western pop.

\subsection{Training Details}

Our framework's two main components, the Extractor and the Decoder, are trained separately.

\short{ \begin{itemize}[noitemsep,nolistsep]}
\full{\begin{itemize}}
    \item \textbf{Extractor}: {We adapt the AMT-APC\cite{amt-apc} architecture. 
Unlike the original, which learns from multiple cover versions of a single song, we use one-to-one pairings. To encourage the model to produce a dense feature map rather than a sparse arrangement, we do not use style vectors and significantly lower the loss sampling parameter ($\theta^{\text{matrix}}$). The model was trained for 10 epochs with a batch size of 2.}
    \item \textbf{Decoder}: {We use a GPT-NeoX\cite{gpt-neox} architecture with 8 Transformer layers, 8 attention heads, and a hidden size of 512, totaling approximately 25.5M parameters. The model was trained on sequences of up to 1,024 tokens. We used the AdamW optimizer\cite{adamw} with a learning rate of $2\times10^{-4}$, scheduled with cosine annealing after a 10 epoch linear warmup. The model was trained for 100 epochs with a batch size of 128.}
\end{itemize}

\subsection{Baselines}\label{sec:baselines}

We compare our model against several APCG baselines: PiCoGen2\cite{picogen2}, AMT-APC\cite{amt-apc}, and Music2MIDI\cite{music2midi}. Furthermore, we include the Human-performed piano cover as an upper bound for comparison. We also evaluate three versions of our own system:
\short{ \begin{itemize}[noitemsep,nolistsep]}
\full{\begin{itemize}}
\item \textbf{Etude Extractor}:  The output of our feature Extractor.
\item \textbf{Etude - Default}: 
The Decoder's output using the default (median) style attributes: (1, 1, 1) for polyphony, rhythmic intensity, and note sustain.
\item \textbf{Etude - Prompted}: The Decoder's output with manually selected style prompts for each song.\footnote{The selected style attributes are based on the author's preference. While our model supports per-measure style injection, for this evaluation, a single set of style vectors was applied to the entire piece.}

\end{itemize}

\subsection{Objective Metrics}

Developing reproducible objective metrics for APCG is crucial. We propose three novel metrics designed to capture key dimensions of quality that correlate with human perception: similarity to the source, rhythmic fluency, and expressive dynamics.

\short{ \begin{itemize}[noitemsep,nolistsep]}
\full{\begin{itemize}}
    \item \textbf{Warp Path Deviation (WPD)}  
 evaluates structural similarity between the generated cover and the original song. Given a DTW warp path ${(S_1,P_1), ..., (S_n,P_n)}$ obtained from chroma features via MrMsDTW\cite{mrmsdtw}, we fit a linear regression $\hat{S_k} = a \cdot P_k + b$ to capture the global tempo trend, then compute WPD as the standard deviation of residuals:
 \full{
\[ \mathrm{WPD} = \sqrt{\frac{1}{L} \sum_{k=1}^{L} \left( S_k - \hat{S_k} \right)^2} \]
}
\short{
 
$\mathrm{WPD} = \sqrt{\frac{1}{L} \sum_{k=1}^{L} \left( S_k - \hat{S_k} \right)^2} $

}
This approach improves upon WP-std from Music2MIDI\cite{music2midi} by remaining robust to global tempo variations while detecting structural misalignments such as missing bars or misinterpreted chords.

    \item \textbf{Rhythmic Grid Coherence (RGC)}
 evaluates rhythmic fluency by quantifying how well inter-onset intervals (IOIs) adhere to a metrical grid. We extract all unique onset times to compute the IOI sequence, identify the most frequent IOIs as the core rhythm set $I$, then search for the optimal base tempo unit $\tau\in I$ that minimizes grid deviation:
 \full{
\[ \mathrm{RGC} = \min_{\tau \in \mathcal{I}} \left( \frac{1}{|\mathcal{I}|} \sum_{i \in \mathcal{I}} \left| \frac{i}{\tau} - \text{round}\left(\frac{i}{\tau}\right) \right| \right)\]
  }
\short{

$\mathrm{RGC} = \min_{\tau \in \mathcal{I}} \left( \frac{1}{|\mathcal{I}|} \sum_{i \in \mathcal{I}} \left| \frac{i}{\tau} - \text{round}\left(\frac{i}{\tau}\right) \right| \right)$

}
A lower RGC indicates more precise alignment to a consistent metrical grid, reflecting higher rhythmic coherence.

   \item \textbf{IOI Pattern Entropy (IPE)} 
assesses the dynamic complexity of rhythmic patterns. We extract the IOI sequence, apply log-transformation and k-means clustering to discretize IOIs into symbols, then compute the Shannon entropy over all overlapping n-grams:
  \full{ \[  \mathrm{IPE} = -\sum_{g \in G_n} P(g) \log_2 P(g) \]
}
\short{

$\mathrm{IPE} = -\sum_{g \in G_n} P(g) \log_2 P(g)$

}
  where $G_n$ is the set of unique n-gram patterns. A higher IPE suggests more varied and less repetitive rhythmic motifs, balancing between monotonous (low IPE) and chaotic (excessively high IPE) patterns.
\end{itemize}

\subsection{Subjective Evaluation}\label{sec:subjective-evaluation}

To complement our objective metrics, we conducted a subjective listening test with 101 volunteers, who were categorized into three groups based on their musical experience: amateurs ($<$3 years), intermediate (3-10 years), and expert ($>$10 years). 
Each participant was asked to listen to four randomly selected songs. For each song, they first heard the original version, followed by seven piano cover versions presented in a randomized and anonymous order.
These versions included the human performance and the outputs from the six models listed in Section \ref{sec:baselines}. To ensure fairness, all model-generated MIDI files and the human performance (which was first transcribed to MIDI) were synthesized into audio using the same piano synthesizer. All audio excerpts were 60-second clips extracted from the beginning of each song.

Following the methodology of PiCoGen2\cite{picogen2}, participants rated each cover on a 5-point Likert scale across four criteria:
\short{ \begin{itemize}[noitemsep,nolistsep]}
\full{\begin{itemize}}
\item \textbf{Similarity (SI)}:  How similar the cover is to the original song in melody, harmony, and rhythm. 
\item \textbf{Fluency (FL)}: The musical coherence and smoothness of the performance.\footnote{A low score may indicate issues such as rhythmic errors, missed notes, or awkward silences that disrupt the musical flow.}
\item \textbf{Dynamic Expression (DE)}: How human-like and expressive the performance sounds.\footnote{A low score may suggest the performance is mechanical, lifeless, or emotionally flat.}
\item \textbf{Overall (OVL)}: How much do the participants like the piano cover in the personal overall listening experience?
\end{itemize}

\subsection{Results and Discussion}

\full{\subsubsection{Evaluation Results}}

The comprehensive results for both objective and subjective evaluations are presented in Table \ref{tab:evaluation}.

The objective metrics reveal that extreme scores are not always optimal; instead, the human performance provides a balanced target. For instance, in WPD, the lowest scores were achieved by models that perform overly literal transcriptions (Etude Extractor, AMT-APC). Similarly, for RGC, the Etude Decoder versions (Etude - Default, Etude - Prompted) achieved the lowest scores, indicating perfect adherence to a metrical grid, whereas the Human performance showed slightly higher deviation due to expressive timing. For IPE, the Human performance struck a balance between the monotonous patterns of PiCoGen2 (low IPE) and the chaotic randomness of AMT-APC (high IPE). This suggests that for objective metrics, closeness to the human benchmark is a strong indicator of a well-balanced and high-quality arrangement.

The subjective evaluation results in Table \ref{tab:evaluation} confirm the superiority of our proposed framework. Statistical analysis reveals that all three Etude variants significantly outperform all three baseline models (PiCoGen2, AMT-APC, and Music2MIDI) in Overall Quality (OVL), with all planned pairwise comparisons yielding a corrected $p<0.001$.

Diving into the details, the Human performance unsurprisingly achieved the highest scores across all criteria, establishing a clear upper bound. Among the AI models, both versions of our Etude Decoder were the top performers, with Etude - Default achieving the highest OVL score of 3.50.
Notably, our decoder versions also lead in Fluency (FL) and Dynamic Expression (DE), indicating that their outputs are perceived as more natural and musical. Interestingly, Etude - Default slightly outperformed Etude - Prompted. This is likely because, for fairness in evaluation, we applied fixed style vectors across all bars in the Prompted setting, which may have limited dynamic expressivity compared to the Default's median attributes.
Meanwhile, the Etude Extractor achieved the highest Similarity (SI) score among all models, confirming its effectiveness at capturing a dense and accurate representation of the source material. These results provide strong evidence that the Etude framework represents a substantial advancement in the quality of automatic piano cover generation.

\full{
\begin{table}[h]
\caption{Descriptive statistics of the Overall (OVL) scores for participant groups based on musical experience. Scores are presented as mean $\pm$ standard deviation.}
\centering
\begin{tabular}{lccc}
\toprule
\textbf{Experience Group} & \textbf{N} & \textbf{OVL Score} \\
\midrule
Amateurs ($<$ 3 years) & 30 & 3.32 $\pm$ 0.53 \\
Intermediate (3-10 years) & 39 & 3.09 $\pm$ 0.50 \\
Expert ($>$ 10 years) & 32 & 3.00 $\pm$ 0.60 \\
\bottomrule
\end{tabular}
\label{tab:experience-group-stats}
\end{table}

\subsubsection{Analysis of Rater Experience}
To validate our subjective findings, we analyzed the OVL scores by rater experience (Table \ref{tab:experience-group-stats}). While a trend emerged where more experienced listeners gave harsher scores, a statistical test showed that the difference among the groups was not significant ($p=0.070$). This suggests our main results are robust across listeners with varying musical expertise.
}

\full{
\subsubsection{Ablation Study}
Our proposed Etude framework introduces significant improvements over prior works in both its data processing pipeline and model architecture. To disentangle these effects and investigate the primary source of the performance gains, we conducted an ablation study focusing on the strongest baseline model, PiCoGen2.

To isolate the impact of the model architecture from the influence of the dataset, we first trained a new version of the PiCoGen2 model, hereafter referred to as PiCoGen2*, following its official implementation but using our own high-quality dataset (detailed in Section \ref{sec:dataset}). We then conducted a separate subjective listening test with 30 new participants. The experimental setup was identical to our main evaluation (Section \ref{sec:subjective-evaluation}), including the audio synthesis process and the four criteria (SI, FL, DE, OVL). The test compared the following three systems:

\begin{itemize}
\item \textbf{Etude - Default (Ours)}: Our proposed model trained on our proposed dataset.
\item \textbf{PiCoGen2* (Ours)}: The baseline model retrained on our proposed dataset.
\item \textbf{PiCoGen2 (Original)}: The official model trained on its original dataset.
\end{itemize}

The results of this ablation study are presented in Table \ref{tab:ablation-study}. The analysis reveals two key findings. First, while retraining the baseline on our dataset (PiCoGen2*) shows a slight improvement over the original PiCoGen2, the difference in OVL scores was not statistically significant ($p=0.0656$). Second, and more crucially, our Etude - Default model significantly outperformed PiCoGen2* ($p=0.0007$), even when both were trained on the exact same dataset. This provides strong evidence that while our data processing pipeline contributes positively, the primary driver for the substantial performance gain is the architectural innovations of the Etude framework itself.

\begin{table}[t]
\caption{Subjective evaluation results for the ablation study, comparing the effects of the dataset and model architecture.}
\centering
\setlength{\tabcolsep}{4pt}
\begin{tabular}{l cccc}
\toprule
\textbf{Model} & \textbf{SI $\uparrow$} & \textbf{FL $\uparrow$} & \textbf{DE $\uparrow$} & \textbf{OVL $\uparrow$} \\
\midrule
Etude - Default (Ours) & \textbf{3.38} & \textbf{3.53} & \textbf{3.58} & \textbf{3.55} \\
PiCoGen2* (Ours) & 3.10 & 3.37 & 3.26 & 3.18 \\
PiCoGen2 (Original) & 2.97 & 3.19 & 2.63 & 2.98 \\
\bottomrule
\end{tabular}
\label{tab:ablation-study}
\end{table}
}

\full{
\section{Conclusion}
In this paper, we addressed the key challenges of structural inconsistency and a lack of stylistic control in Automatic Piano Cover Generation (APCG). We proposed \textbf{Etude}, a novel three-stage framework built upon a rigorous data processing pipeline. Our approach modularizes the task into three distinct components—a Beat-Detector, an Extractor, and a Decoder—and introduces \textbf{Tiny-REMI}, a minimalist and efficient token representation. This decoupled design simplifies the learning process and, by leveraging a pre-extracted rhythmic framework, guarantees that the generated output is structurally consistent with the original song. Furthermore, by introducing a mechanism for injecting relative style attributes, our framework enables the generation of stylistically diverse and controllable piano covers.

Our main contributions include a highly extensible modular framework, an efficient Tiny-REMI representation that simplifies the learning task, and a novel style injection mechanism for controllable generation. Through extensive subjective evaluations, our Etude framework was shown to significantly outperform all state-of-the-art baselines, achieving a quality that approaches that of human arrangers.

For future work, we observe that
the performance of our framework is fundamentally upper-bounded by the capabilities of its front-end components: the Beat-Detector and the Extractor. The framework's structural accuracy is limited by the precision of the beat tracker, while the Extractor's process of flattening the source audio into a single feature stream can lead to information loss. This limitation is evident in our subjective results (Table \ref{tab:evaluation}), where both Etude Decoder versions scored markedly lower on Similarity (SI) than the direct Etude Extractor output. 
We hypothesize this is because the flattened features make it difficult for the model to identify the primary melody of the original song, causing the Decoder to generate incomplete melodic lines. Future work could thus explore both the integration of more advanced beat-tracking modules and the development of multi-stream extractors capable of disentangling different musical sources. Finally, our work highlights the persistent gap between current objective metrics and human perception. Developing learnable, perception-aligned evaluation models could provide a more reliable and automated benchmark for the field. Overall, the Etude framework provides a robust foundation for generating high-quality, controllable piano covers, paving the way for more musical and interactive AI-powered creative tools.
}

\hide{
\section*{Compliance with Ethical Standards statement}
\begin{itemize}
\item No funding was received for conducting this study. The authors have no relevant financial or nonfinancial interests to disclose.
\item To our knowledge, no ethical approval was required for this  study.
\end{itemize}
}

\bibliographystyle{IEEEbib}
\short{\small{
\bibliography{ICASSP2026_ref}
}
}
\full{
\bibliography{ICASSP2026_ref}
}
\end{document}